# An Alignment Microscope for Imaging Rotational Wavepacket Dynamics

J. STROHABER,[1*]

[1] *Department of Physics, Florida A&M University, Tallahassee, Florida 32307, USA*
**james.strohaber@famu.edu*

**Abstract:** We introduce a new method for encoding polarization-angle-dependent molecular alignment signals directly into the transverse spatial profile of a femtosecond laser beam. The approach employs a pump-probe setup in a common-path geometry. A radial–azimuthal polarization converter placed in the probe beam produces a doughnut-shaped beam whose local polarization varies continuously with azimuthal angle, thereby encoding the full range of linear polarization orientations from 0 to $2\pi$ within a single transverse beam profile. Consequently, the angular dependence of the molecular response can be recorded simultaneously rather than reconstructed through sequential measurements at different polarization angles. By varying only the pump–probe delay, this spatial encoding enables single-shot acquisition of the complete angular response at each delay and real-time visualization of the evolving molecular alignment pattern as the delay stage is scanned. This technique has enabled the experimental realization of an alignment microscope.

## 1. Introduction

The interaction of intense femtosecond laser pulses with molecules provides opportunities for investigating and controlling molecular dynamics on ultrafast timescales. Laser-induced molecular alignment has attracted considerable interest because it enables control over the angular distribution of molecular axes through their anisotropic polarizability [1–4]. When a nonresonant, linearly polarized laser pulse interacts with a molecule, the induced dipole interaction produces a torque that preferentially aligns the molecular axis along the laser polarization direction. For excitation pulses substantially shorter than the characteristic rotational period, the interaction generates a coherent superposition of rotational states that evolves freely after the pulse has passed. The resulting rotational wave packet undergoes periodic dephasing and rephasing, producing transient alignment and antialignment at characteristic revival times determined by the rotational energy-level structure [2–5].

Field-free molecular alignment has been investigated extensively in coherent control, molecular imaging, and nonlinear optics. Experimental studies have demonstrated alignment at room temperature and enhanced alignment using appropriately delayed excitation pulses [6], while theoretical investigations have established methods for controlling rotational dynamics through spectral phase shaping [7,8]. Molecular alignment has also been employed in studies of strong-field ionization, rotational dynamics, and high-order harmonic generation [9–11]. In gaseous media, coherent rotational motion produces transient changes in the refractive index and birefringence that influence ultrashort-pulse propagation. Bartels et al. demonstrated alignment-induced phase modulation and spectral broadening of a delayed probe pulse [12], while subsequent studies investigated molecular alignment in femtosecond filamentation and nonlinear optical propagation [13,14]. Collisional relaxation of rotational coherence has also been examined theoretically and experimentally [15,16].

Several experimental techniques have been developed to characterize rotational wave-packet dynamics. Coulomb-explosion imaging and ion detection provide measurements of molecular angular distributions and fractional rotational revivals [17,18]. Complementary optical methods detect alignment-induced changes in the refractive index, polarization, and spectral properties of a delayed probe. Feng et al. demonstrated detection of field-free alignment through spatial focusing and defocusing [19], while Houzet et al. employed coherent 4*f* imaging to distinguish alignment and antialignment through phase-sensitive measurements [20]. Polarization-resolved birefringence measurements have also been used to characterize planar molecular alignment [21]. Additionally, spectral modulation of a delayed probe has enabled reconstruction of rotational wave-packet dynamics, including angularly resolved quantum carpets obtained by varying the relative pump and probe polarization directions [22]. Although these techniques are sensitive to molecular alignment, angularly resolved measurements generally require sequential changes in the probe polarization.

Spatially structured optical fields offer an alternative approach. Cylindrical vector beams, including radially and azimuthally polarized beams, possess spatially varying polarization distributions that encode multiple polarization orientations within a single transverse beam profile [23]. When such a field probes impulsively aligned molecules, the local polarization direction determines its interaction with the anisotropic molecular response. Consequently, the transmitted intensity distribution can reveal angularly dependent optical signatures of molecular alignment without sequential polarization measurements. These signatures may arise from several mechanisms, including alignment-induced birefringence, absorption, and cross-phase modulation, whose relative contributions depend on the polarization geometry and spectral detection conditions.

Because molecular dimensions are far smaller than optical wavelengths, individual molecules cannot be spatially resolved using conventional optical microscopy. Nevertheless, the coherent quantum-mechanical evolution of rotational wave packets provides a means of visualizing molecular dynamics through the collective optical response of an ensemble. In this work, we exploit this principle to demonstrate an *alignment microscope* for spatially and angularly resolved imaging of rotational wave-packet dynamics in impulsively aligned nitrogen molecules. Using femtosecond pump–probe spectroscopy, spatially varying probe polarization, and CMOS detection, multiple polarization orientations are measured simultaneously at each pump–probe delay. Polarization-resolved detection, integrated power measurements, spectral analysis, and spatial imaging are employed to investigate the contributions of birefringence, absorption, and molecular-alignment-induced cross-phase modulation (MA-XPM). These mechanisms produce optical signatures associated with the molecular alignment parameter, $\Delta\langle\cos^2\theta\rangle$, and its temporal derivative, $d\langle\cos^2\theta\rangle/d\tau$. The resulting approach enables visualization of collective rotational dynamics through spatially and angularly resolved optical measurements.

## 2. Theoretical Treatment

The theoretical description of field-free molecular alignment has been extensively developed in the literature [1—5]. Here, we briefly review the fundamental principles of nonadiabatic rotational alignment in diatomic molecules, emphasizing the formation and evolution of rotational wave packets following excitation by an ultrashort laser pulse. We further consider alignment-dependent absorption, birefringence, and molecular-alignment-induced cross-self-phase modulation (MA-XSPM) as possible mechanisms contributing to the measured optical response. Particular attention is given to identifying observables associated with the degree of molecular alignment and its temporal derivative. These considerations provide the theoretical foundation for interpreting the experimental measurements and establishing a relationship between rotational wave-packet dynamics and the spatially resolved optical signals recorded by the alignment microscope.

### *2.1 Alignment Signal*

We consider a homonuclear diatomic molecule interacting with a linearly polarized femtosecond laser pulse. Assuming a Gaussian temporal intensity profile, the complex electric field can be expressed as

$$F(t) = F_0(\mathbf{r}) \exp\left[-2\ln(2)\frac{t^2}{\tau_p^2}\right] e^{-i\omega t}\,, \tag{1}$$

where $F_0(\mathbf{r})$ is the spatially dependent electric-field amplitude, $\tau_p$ is the full width at half maximum (FWHM) of the intensity envelope, and $\omega$ is the optical angular frequency. Within the rigid-rotor approximation, the time-dependent Schrödinger equation governing the rotational dynamics is given by Eq. (2) [1—4].

$$i\hbar\frac{d}{dt}\Phi_{JM}(t) = \left[H_R^{(0)} + V_L(t)\right]\Phi_{JM}(t)\,, \tag{2}$$

where $H_R^{(0)}$ is the field-free rotational Hamiltonian and $V_L(t)$ describes the interaction between the induced molecular dipole and the laser field. For a nonresonant, linearly polarized laser pulse, the cycle-averaged interaction potential is given by [1—4]

$$V_L(t) = -\frac{1}{4}(\Delta\alpha\cos^2\theta + \alpha_\perp)\left|F_0^2(\mathbf{r})\right|^2 \exp\left[-4\ln(2)\frac{t^2}{\tau_p^2}\right]. \tag{3}$$

The molecular polarizability anisotropy is $\Delta\alpha = \alpha_\parallel - \alpha_\perp$, where $\alpha_\parallel$ and $\alpha_\perp$ are the polarizabilities parallel and perpendicular to the molecular axis, respectively. The angle $\theta$ is measured between the molecular axis and the laser polarization direction. The anisotropic interaction produces a coherent superposition of rotational states, forming a rotational wave-packet that evolves after the excitation pulse has passed.

To determine the time-dependent molecular alignment, the rotational wavefunction is expanded in terms of the field-free rotational eigenstates,

$$\left|\Phi_{J_0M_0}(t)\right\rangle = \sum_{JM} d_{JM}^{J_0M_0}(t)\exp\left(-i\frac{E_J}{\hbar}t\right)\left|JM\right\rangle. \tag{4}$$

Here, $J$ is the rotational angular momentum quantum number and $M$ is its projection along the laser polarization axis. The initial rotational state is specified by $J_0$ and $M_0$, while $d_{JM}^{J_0M_0}(t)$ represents the time-dependent expansion coefficients. For a rigid rotor, the rotational energy eigenvalues are $E_J = BJ(J+1)$ where $B = \hbar^2/2I$ is the rotational constant expressed in energy units and $I$ is the molecular moment of inertia.

The degree of molecular alignment is characterized by the expectation value

$$\left\langle\cos^2(\theta)\right\rangle_{J_0M_0}(t) = \left\langle\Phi_{J_0M_0}(t)\right|\cos^2(\theta)\left|\Phi_{J_0M_0}(t)\right\rangle. \tag{5}$$

For an isotropic molecular ensemble, this ensemble-averaged alignment parameter equals 1/3. Values greater than 1/3 indicate preferential alignment along the laser polarization direction, whereas values below 1/3 indicate preferential alignment perpendicular to that direction. Following impulsive excitation, the coherent rotational wave packet undergoes periodic rephasing, producing characteristic alignment revivals [2–4].

For molecular nitrogen, the rotational constant is approximately $B = \hbar^2/2I \approx 3.97\times10^{-23}\,J$ corresponding to $B \approx 0.248$ meV. At room temperature the thermal energy is approximately $k_BT = 25$ meV, considerably larger than the energy separation between the lowest rotational levels. Consequently, a broad distribution of rotational states is initially populated, and the observed alignment response must account for the thermal population of the molecular ensemble. The initial population of a rotational level $J_0$ is determined by the Boltzmann distribution, $P_J = g_{J_0}(2J+1)\exp(-E_J / k_B T)/Z$ where $g_{J_0}$ is the nuclear-spin statistical weight and $Z$ is the rotational partition function [3,4]. The ensemble-averaged alignment at temperature $T$ is therefore

$$\left\langle\left\langle \cos^2(\theta)\right\rangle\right\rangle(\tau) = \frac{\sum_{J_0}\sum_{M_0=-J_0}^{J_0} g_{J_0}\left\langle \cos^2(\theta)\right\rangle_{J_0 M_0}(\tau)\exp\left[-BJ_0(J_0+1)/kT\right]}{\sum_{J_0}\sum_{M_0=-J_0}^{J_0} g_{J_0}\exp\left[-BJ_0(J_0+1)/kT\right]} . \tag{6}$$

For $^{14}N_2$, nuclear-spin statistics give relative weights of 6 and 3 for even and odd rotational states, respectively, resulting in an even-to-odd statistical weight ratio of 2:1. Finally, the alignment response relative to an isotropic molecular distribution is defined as $\Delta\langle\cos^2\theta\rangle(\tau) = \langle\cos^2\theta\rangle(\tau) - 1/3$ where $\tau$ denotes the pump–probe delay. This quantity describes the time-dependent departure from isotropic molecular alignment and serves as the theoretical reference for interpreting the experimentally measured optical signals. Throughout the remainder of this work, we define $A(\tau) = \langle\cos^2\theta\rangle(\tau)$. Therefore, $\Delta A(\tau) = A(\tau) - 1/3$ and its temporal derivative, $dA/d\tau$, are subsequently used to investigate the alignment-dependent intensity and spectral modulations observed in the experiment.

*2.2 Absorption and Birefringence*

We examine the optical response of aligned molecules in terms of the complex refractive index. Its real component governs alignment-induced birefringence and phase modulation, while its imaginary component describes optical absorption. Molecular alignment modifies the effective polarizability and, consequently, the refractive index experienced by the probe pulse [12,19,20]. These changes provide a connection between the theoretical degree of molecular alignment and the measured optical response, including the horizontal and vertical polarization components detected using a polarizing beam splitter (PBS).

The complex refractive index is $\tilde{n} = n + i\kappa$ where $n$ is the real refractive index and $\kappa$ is the extinction coefficient. In this manuscript, the pump field is always horizontally polarized, and the subscripts $H$ and $V$ denote probe polarization components parallel and perpendicular to the pump polarization direction, while the subscripts $\|$ and $\perp$ reserved for quantities parallel and perpendicular to the molecular axis respectively. For a molecular ensemble aligned along the pump polarization, the effective complex polarizability experienced by the horizontally polarizing probe component is $\tilde{\alpha}_H(\tau) = \tilde{\alpha}_\perp + (\tilde{\alpha}_\| - \tilde{\alpha}_\perp)A(\tau)$ . Here, $\tilde{\alpha}_\|$ and $\tilde{\alpha}_\perp$ are the molecular polarizabilities parallel and perpendicular to the molecular axis. For a dilute gas, the complex refractive index is approximately $\tilde{n}_H(\tau) \approx 1 + N\tilde{\alpha}_H(\tau)/2\varepsilon_0$ where $N$ is the molecular number density and $\varepsilon_0$ is the vacuum permittivity. Using $\Delta A(\tau) = A(\tau) - 1/3$ and $\Delta\tilde{\alpha} = \tilde{\alpha}_\| - \tilde{\alpha}_\perp$, the alignment-induced refractive-index change becomes

$$\Delta\tilde{n}_H(\tau) \approx \frac{N\Delta\tilde{\alpha}}{2\varepsilon_0}\Delta A(\tau) . \tag{7}$$

This expression establishes the relationship between molecular alignment and the complex refractive index. Separating the real and imaginary components, $\Delta\tilde{\alpha} = \Delta\tilde{\alpha}' + i\Delta\tilde{\alpha}''$, gives

$$\Delta n_H(\tau) \approx \frac{N\Delta\tilde{\alpha}'}{2\varepsilon_0}\Delta A(\tau) , \tag{8a}$$

$$\Delta\kappa_H(\tau) \approx \frac{N\Delta\tilde{\alpha}''}{2\varepsilon_0}\Delta A(\tau) . \tag{8b}$$

Equation (8a) describes the alignment-dependent refractive-index change responsible for optical phase modulation, while Eq. (8b) describes the corresponding change in the extinction coefficient. Importantly, a nonzero real polarizability anisotropy does not imply that the imaginary polarizability anisotropy is appreciable. For propagation through a gas of length $L$, the transmitted horizontal electric field with the complex phase $\varphi_H = k_0 \tilde{n}_H L$ is

$$E_{H,\text{out}}(\tau) = E_{H,\text{in}} e^{ik_0 n_H L} e^{-k_0 \kappa_H L} , \tag{9}$$

where $k_0 = 2\pi / \lambda_0$ is the vacuum wave number. The optical power is the modulus squared of the field $P_H(\tau) = P_{\text{in}} \exp(-2k_0\kappa_H L)$. Writing $\kappa_H = \kappa_{H0} + \Delta\kappa_H$ where $\kappa_{H0}$ is the extinction coefficient for an isotropic molecular ensemble gives $P_H(\tau) = P_{\text{in}} \exp[-2k_0(\kappa_{H0} + \Delta\kappa_H)L]$. Separating the baseline attenuation this can be written as $P_H(\tau) = P_{H0} \exp(-2k_0 \Delta\kappa_H L)$, we obtain

$$P_H(\tau) = P_{H0} \exp\left[-k_0 L \frac{N}{\varepsilon_0} \Delta\alpha'' \Delta A(\tau)\right] . \tag{10}$$

The fractional power change is therefore

$$\frac{\Delta P_H}{P_{H0}} = \frac{P_H - P_{H0}}{P_{H0}} = \exp\left[-k_0 L \frac{N}{\varepsilon_0} \Delta\alpha'' \Delta A(\tau)\right] - 1 . \tag{11}$$

For weak alignment-dependent absorption, we can expand the exponential: $e^{-x} \approx 1 - x$, yielding

$$\frac{\Delta P_H}{P_{H0}} \approx -\frac{k_0 L N}{\varepsilon_0} \Delta\alpha'' \Delta A(\tau) . \tag{12}$$

For the vertically polarized probe component $\tilde{\alpha}_V(\tau) = \tilde{\alpha}_\perp + \Delta\tilde{\alpha}[1 - A(\tau)]/2$ assuming cylindrical symmetry of the aligned molecular ensemble. Consequently, the alignment-induced changes in the vertical extinction coefficient and horizontal extinction coefficient are,

$$\Delta\kappa_V(\tau) \approx -\frac{N\Delta\alpha''}{4\varepsilon_0}\Delta A(\tau) , \tag{13a}$$

$$\Delta\kappa_H(\tau) \approx \frac{N\Delta\alpha''}{2\varepsilon_0}\Delta A(\tau) . \tag{13b}$$

The corresponding weak-absorption signal for the vertical component is

$$\frac{\Delta P_V}{P_{V0}} \approx +\frac{k_0 L N}{2\varepsilon_0} \Delta\alpha'' \Delta A(\tau) . \tag{14}$$

These expressions predict opposite signs for the alignment-dependent attenuation of the horizontal and vertical probe components when the imaginary polarizability anisotropy is nonzero. However, the relative amplitudes depend on the optical properties and detection geometry, and the measured polarization-resolved intensity changes may also contain contributions from birefringence and spatial or spectral redistribution [19,20,22].

*2.3 Molecular-Alignment Cross-Self-Phase Modulation* (*MA-XSPM*)

The optical Kerr effect gives rise to an intensity-dependent refractive index that produces self-focusing and self-phase modulation (SPM), leading to modifications of the optical spectrum. Previous studies have shown that molecular alignment can influence these nonlinear optical processes, including self-focusing, filamentation, and spectral broadening [13]. In conventional SPM, the optical pulse modulates its own phase through the intensity-dependent refractive index [24]. Here, we consider an effective nonlinear refractive index that depends on both the instantaneous probe intensity and the molecular alignment induced by the pump pulse. We refer to this combined process as molecular-alignment cross–self-phase modulation (MA-XSPM) [25–27].

The conventional Kerr expression for the refractive index is $n(t) = n_0 + n_{\mathrm{NL}} I(t)$ where $n_0$ is the linear refractive index, $n_{\mathrm{NL}}$ is the nonlinear refractive-index coefficient, and $I(t)$ is the optical intensity. The pump pulse induces rotational molecular alignment that evolves on a picosecond timescale, whereas the probe intensity varies on a femtosecond timescale. At a pump-probe delay $\tau$, the effective nonlinear coefficient is modeled as $n_{\mathrm{NL}}(t,\tau) = n_{\mathrm{NL,iso}} + \eta \Delta A(t+\tau)$ where $\Delta A = A - 1/3$ *and* $\eta$ describes the alignment-dependent modifications of the nonlinear response. The accumulated nonlinear phase is $\phi_{\mathrm{NL}}(t,\tau) = k_0 L n_{\mathrm{NL}}(t,\tau) I(t)$ with $k_0 = 2\pi/\lambda_0$. The nonlinear phase separates into a background self-phase modulation (SPM) contribution and an alignment-dependent contribution arising from the pump-induced molecular response,

$$\phi_{\mathrm{NL}}(t,\tau) = k_0 L n_{\mathrm{NL,iso}} I(t) + \eta k_0 L \Delta A(t+\tau) I(t) . \tag{15}$$

Because the probe duration is short compared with the characteristic variation time of molecular alignment, a first-order Taylor expansion gives $\Delta A(t+\tau) \approx \Delta A(\tau) + tA'(\tau)$. Writing $I(t) = I_0 f(t)$, where $f(t)$ is the normalized intensity profile, and defining $B(\tau) = k_0 L I_0 [n_{\mathrm{NL,iso}} + \eta \Delta A(\tau)]$ and $D(\tau) = \eta k_0 I_0 L A'(\tau)$, the nonlinear phase becomes $\phi_{\mathrm{NL}}(t,\tau) = [B(\tau) + D(\tau)t] f(t)$.

For convenience, we introduce the pulse-shape coefficients $K_0, K_1, K_2$, and $K_3$, defined below for a dimensionless intensity profile $f(t)$, normalized to unity at its peak, with $\int_{-\infty}^{\infty} f(t)dt = T_f$.

$$K_0 = \frac{1}{T_f} \int_{-\infty}^{\infty} f^2(t) dt , \tag{16a}$$

$$K_1 = \frac{1}{T_f} \int_{-\infty}^{\infty} f(t)[f'(t)]^2 dt , \tag{16b}$$

$$K_2 = \frac{1}{T_f} \int_{-\infty}^{\infty} f(t) f'(t)[f(t) + tf'(t)] dt , \tag{16c}$$

$$K_3 = \frac{1}{T_f} \int_{-\infty}^{\infty} f(t)[f(t) + tf'(t)]^2 dt - K_0^2/4 . \tag{16d}$$

The instantaneous frequency shift is $\Delta\omega(t,\tau) = -\partial \phi_{NL} / \partial t$. Differentiating the nonlinear phase gives $\Delta\omega(t,\tau) = -Bf'(t) - D[f(t) + tf'(t)]$. The spectral centroid shift is obtained by averaging the instantaneous frequency shift over the normalized probe intensity profile,

$$\Delta\bar{\omega} = \frac{1}{T_f}\int f(t)\Delta\omega(t,\tau)dt\,. \tag{17}$$

Because the intensity vanishes at the integration boundaries, $\int f(t)f'(t)dt = 0$. Integrating by parts also gives $\int tf(t)f'(t)dt = -\int f^2(t)dt/2$. Using the definition of $K_0$, the spectral centroid shift becomes $\Delta\bar{\omega}(\tau) = -K_0 D(\tau)/2$. Converting from angular frequency to wavelength using $ck = \omega$ and $\Delta\lambda_c \approx (\partial\lambda/\partial\omega)_{\omega_0}\Delta\omega$, we obtain

$$\Delta\lambda_c(\tau) = \frac{\lambda_0 L I_0 \eta K_0}{2c}\frac{dA}{d\tau}\,. \tag{18}$$

Thus, the center-wavelength shift is proportional to the time derivative of molecular alignment.

Next, we consider the RMS spectral bandwidth of an initially transform-limited probe with a real electric field envelope $a(t) = \sqrt{I_0 f(t)}$. The spectral variance is given by

$$\sigma_\omega^2(\tau) = \sigma_{\omega,0}^2 + \left\langle[\Delta\omega(t,\tau)]^2\right\rangle_I - \left\langle\Delta\omega(t,\tau)\right\rangle_I^2, \tag{19}$$

where $\sigma_{\omega,0}$ is the initial RMS spectral bandwidth and $\langle\ \rangle_I$ denotes the intensity-weighted average defined previously. For an initially transform-limited probe, substituting the instantaneous frequency shift $\Delta\omega(t,\tau) = -Bf'(t) - D[f(t) + tf'(t)]$ and using the pulse-shape coefficients defined in Eqs. (16a–16d), the spectral variance becomes

$$\sigma_\omega^2(\tau) = \sigma_{\omega,0}^2 + K_1 B^2(\tau) + 2K_2 B(\tau)D(\tau) + K_3 D^2(\tau)\,. \tag{20}$$

Here $K_1$ characterizes the SPM-induced spectral broadening, $K_2$ represents the cross term, and $K_3$ describes the derivative-squared contribution to spectral broadening. These coefficients depend only on the temporal intensity profile of the probe pulse. For a temporally symmetric pulse centered at $t = 0$, the cross coefficient $K_2$ vanishes by symmetry, yielding

$$\sigma_\omega^2(\tau) = \sigma_{\omega,0}^2 + K_1 B^2(\tau) + K_3 D^2(\tau)\,. \tag{21}$$

This result shows that the spectral bandwidth variance depends on both molecular alignment and the square of its time derivative. The alignment-dependent nonlinear phase coefficient is $B^2(\tau) = (k_0 L I_0)^2[n_{\mathrm{NL,iso}} + \eta\Delta A(\tau)]^2$. When the alignment-induced correction is small compared with the isotropic nonlinear coefficient, $|\eta\Delta A| \ll |n_{\mathrm{NL,iso}}|$, a first-order expansion gives $B^2(\tau) \approx (k_0 L I_0)^2[n_{\mathrm{NL,iso}}^2 + 2n_{\mathrm{NL,iso}}\eta\Delta A(\tau)]$. If the derivative-squared contribution is also negligible compared with the alignment-dependent linear contribution, the change in spectral variance can be linearized about the isotropic-background value. For a sufficiently narrow spectrum, linearizing the frequency-to-wavelength conversion about the carrier wavelength $\lambda_0$ gives $\Delta\sigma_\lambda(\tau) \approx (\lambda_0^2/2\pi c)\Delta\sigma_\omega(\tau)$. We therefore obtain

$$\Delta\sigma_\lambda(\tau) \approx \frac{K_1\lambda_0^2(LI_0)^2 n_{\mathrm{NL},iso}\eta}{c^2\sigma_{\lambda,\mathrm{bg}}}\Delta A(\tau)\,, \tag{22}$$

where $\sigma_{\lambda,\mathrm{bg}}$ is the RMS wavelength bandwidth in the isotropic-background limit. Equation (22) predicts that the RMS wavelength bandwidth varies approximately linearly with molecular alignment, whereas Eq. (18) predicts that the center-wavelength shift follows its time derivative. These distinct spectral signatures provide the theoretical basis for interpreting the experimental measurements presented in Sec. 4.3.

In the presence of absorption and spectrally filtered MA-XSPM, the measured intensity modulation can be approximated phenomenologically as

$$\frac{\Delta P}{P_0} = C_{\text{loss}} \Delta A + C_{\text{XSPM}} \frac{dA}{d\tau}, \tag{23}$$

where $C_{\text{loss}}$ and $C_{\text{XSPM}}$ are effective coefficients describing the alignment-dependent and derivative-dependent contributions to the measured intensity modulation, respectively. The relative contributions of these mechanisms may vary across the transverse beam profile because of spatial and angular chirp, local intensity variations, and spatially dependent nonlinear phase accumulation.

### *2.4 Radially and Azimuthally Polarized Beams*

Radially and azimuthally polarized beams are cylindrical vector beams whose local electric-field polarization varies across the transverse beam profile. These fields can be constructed through the coherent superposition of orthogonally polarized spatial modes [28–30]. The following treatment follows Strohaber [28], using even and odd Laguerre–Gaussian modes to describe the transverse and longitudinal electric-field components.

$$\text{LG}_{\rho,\ell}^{(e,o)} = E_0 \frac{w_0}{w} \left( \frac{\sqrt{2} r}{w} \right)^{|\ell|} LG_{\rho}^{|\ell|} \left( \frac{2r^2}{w^2} \right) \exp\left( -\frac{r^2}{w^2} + i \frac{kr^2}{2R} \right) e^{-i(2\rho + |\ell| + 1)\Psi_G} e^{ikz} \begin{bmatrix} \cos(\ell\theta) \\ \sin(\ell\theta) \end{bmatrix}. \tag{24}$$

For $\rho = 0$ and $\ell = 1$ Eq. (24) reduces to the first-order even and odd Laguerre–Gaussian modes, which exhibit doughnut-shaped intensity profiles when combined to form cylindrical vector beams.

$$\text{LG}_{0,1}^{(e,o)} = E_0 \frac{w_0}{w} \left( \frac{\sqrt{2} r}{w} \right) \exp\left( -\frac{r^2}{w^2} + i \frac{kr^2}{2R} \right) e^{-i2\Psi_G} e^{ikz} \begin{bmatrix} \cos(\theta) \\ \sin(\theta) \end{bmatrix}. \tag{25}$$

The radially and azimuthally polarized vector fields are constructed by combining these modes with orthogonal linear polarizations [28,29]: $\vec{E}_R = \text{LG}_{0,1}^{e} \hat{e}_x + \text{LG}_{0,1}^{o} \hat{e}_y$ and $\vec{E}_A = -\text{LG}_{0,1}^{o} \hat{e}_x + \text{LG}_{0,1}^{e} \hat{e}_y$. Here $w(z) = w_0 (1 + z^2 / z_0^2)^{1/2}$ is the beam radius, $w_0$ is the beam waist, $\Psi_G = \arctan(z / z_0)$ is the Gouy phase, and $R(z) = z + z_R^2 / z$ is the radius of curvature. The even and odd modes can be expressed as as $\psi_e = A(r,z)\cos(\phi)$ and $\psi_o = A(r,z)\sin(\phi)$ respectively. Within the paraxial approximation, Maxwell's equations give the electric field as [28]

$$\vec{E}(r,\phi) = E_0 \left[ -i\psi \left( \alpha \hat{e}_x + \beta \hat{e}_y \right) + \frac{1}{k} \left( \alpha \frac{\partial \psi}{\partial x} + \beta \frac{\partial \psi}{\partial y} \right) \hat{e}_z \right]. \tag{26}$$

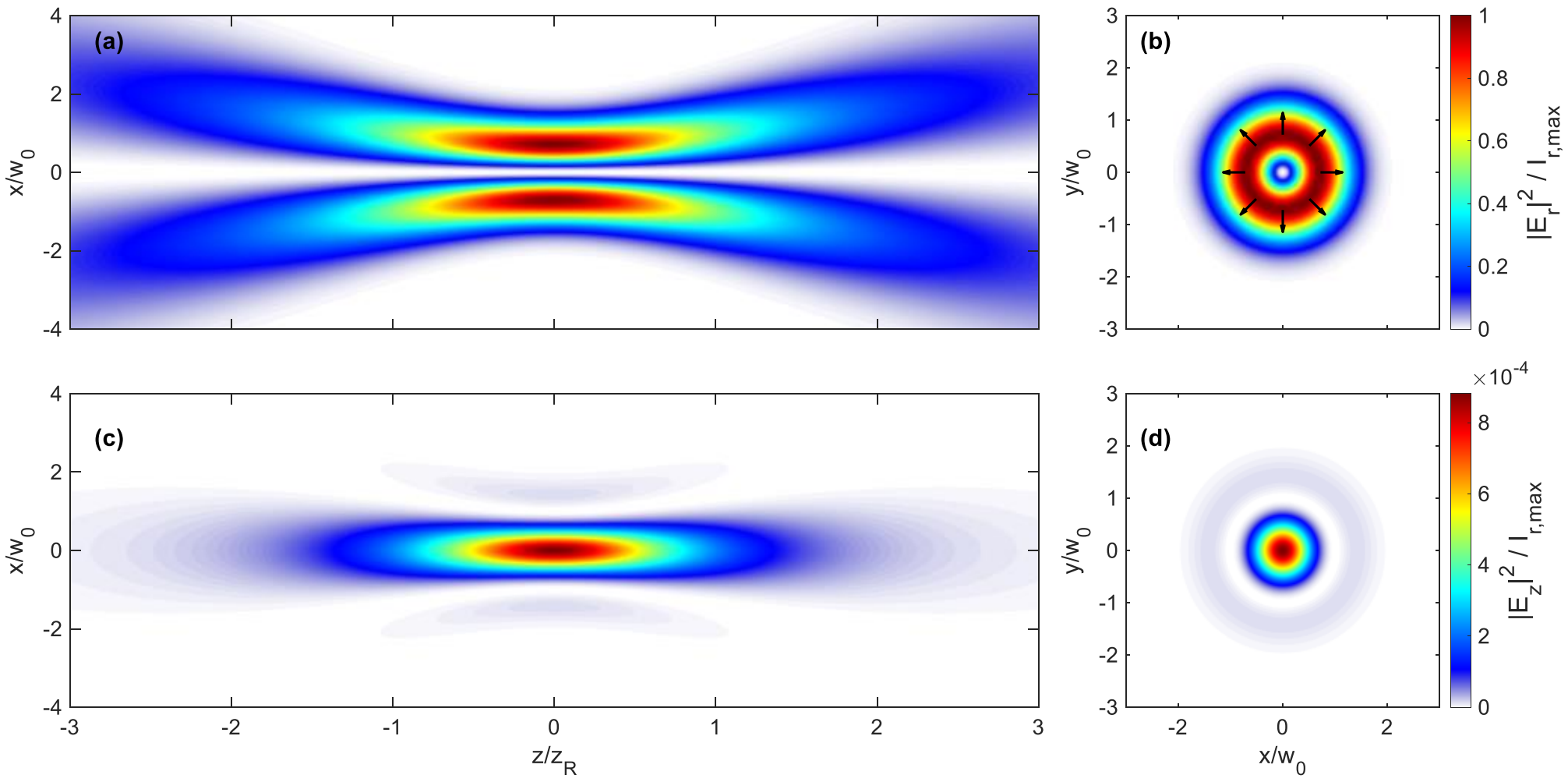

Figure 1: Theoretical spatial intensity distributions of the radial and longitudinal electric-field components of a radially polarized beam propagating along the z-axis. (a) Intensity distribution of the radial electric-field component, $|E_r|^2$, in the *xz*-plane, illustrating beam propagation and the absence of a radial field component along the optical axis. (b) Transverse intensity distribution of the radial component in the focal (*xy*) plane, exhibiting a characteristic doughnut-shaped profile with a central intensity minimum. Arrows indicate the local direction of radial polarization. (c) Intensity distribution of the longitudinal electric-field component, $|E_z|^2$, in the *xz*-plane, showing a maximum along the optical axis near the beam waist. (d) Corresponding transverse intensity distribution of the longitudinal component in the focal plane, exhibiting a centrally localized intensity maximum. All intensities are normalized to the peak radial intensity, allowing direct comparison of the two components. Under the focusing conditions considered, the peak longitudinal intensity is approximately four orders of magnitude smaller than the peak radial intensity. In contrast, an ideal azimuthally polarized beam has no longitudinal electric-field component, and its transverse polarization is everywhere perpendicular to that of a radially polarized beam.

For horizontally polarized light $\alpha = 1$ and $\beta = 0$ . To find the radially and azimuthally polarized fields we use $\hat{e}_r = \cos(\phi)\hat{e}_x + \sin(\phi)\hat{e}_y$ , and $\hat{e}_\phi = -\sin(\phi)\hat{e}_x + \cos(\phi)\hat{e}_y$ along with the even and odd beam modes $\psi_e$ and $\psi_o$ . The resulting fields are

$$\vec{E}_A(r,\phi) = -iE_0 A(r,z)\hat{e}_\phi , \tag{27a}$$

$$\vec{E}_R(r,\phi) = E_0\left[-iA(r,z)\hat{e}_r + \frac{1}{k}\left(\frac{\partial A}{\partial r} + \frac{A}{r}\right)\hat{e}_z\right]. \tag{27b}$$

Two important observations follow from Eqs. (27a) and (27b). First, the ideal azimuthally polarized beam has no longitudinal electric-field component, whereas the radially polarized beam possesses both transverse and longitudinal components [28,31]. The longitudinal component depends on the radial variation of the field amplitude and becomes particularly important under tight focusing. Because the alignment microscope measures a response accumulated over a finite interaction length, the purely transverse polarization of the azimuthally polarized beam is advantageous for the present configuration.

Figure 1 illustrates the transverse and longitudinal intensity distributions of a radially polarized beam. Panel (a) shows the transverse radial component, which retains a doughnut-shaped intensity profile, while panel (b) shows the longitudinal component concentrated near the focal region. The azimuthally polarized beam has the same transverse intensity distribution as the radial component of the radially polarized beam and is therefore not shown.

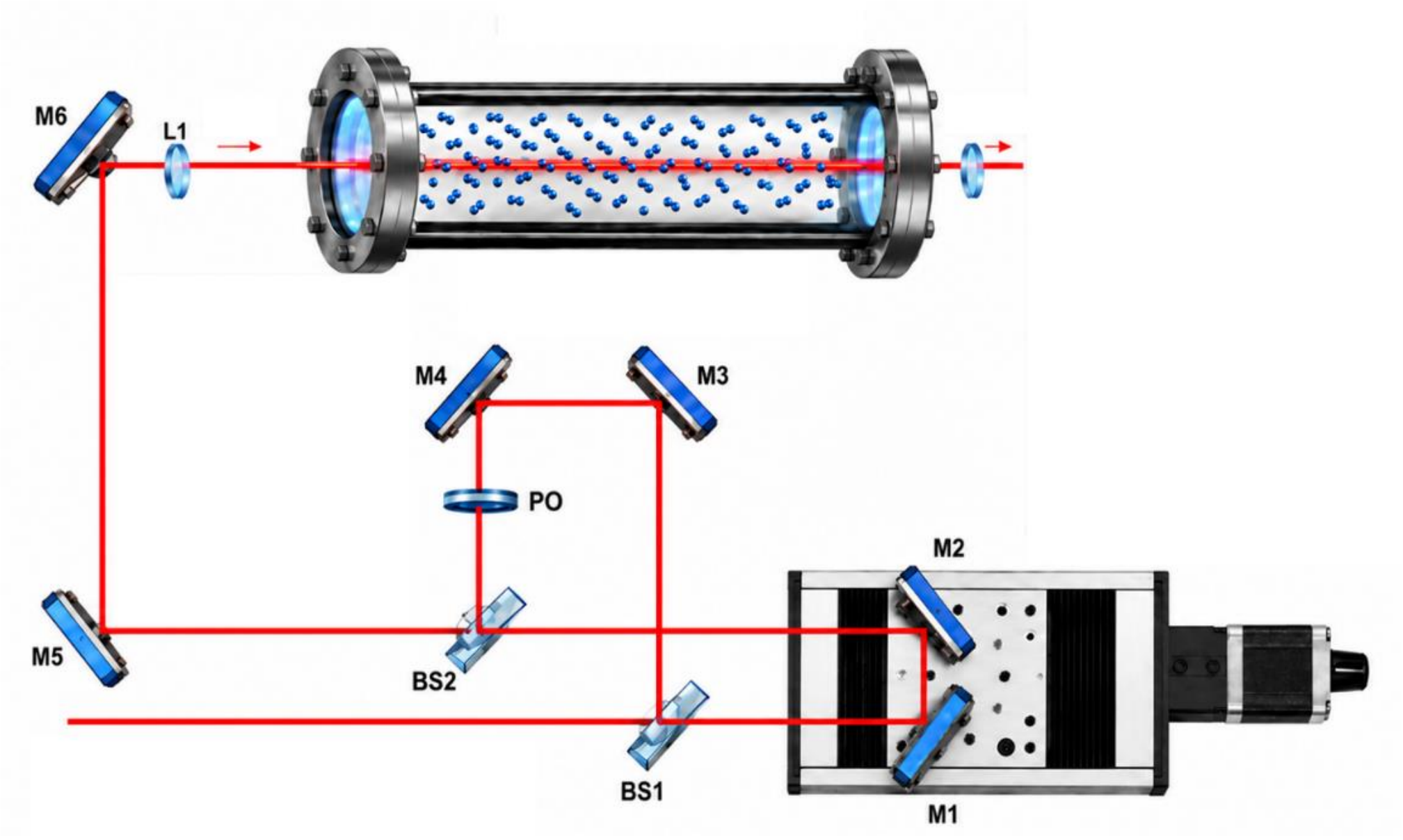

Figure 2: Experimental setup. An 85-cm-long alignment tube contains $N_2$ gas at approximately 4 bar. Femtosecond laser pulses from a chirped-pulse amplification (CPA) system enter a four-port Michelson interferometer consisting of two nonpolarizing beam splitters (BS1 and BS2). Polarization optics (PO) in the stationary arm consist of either a half-wave plate or a radial/azimuthal polarization converter, while a Newport MTM150PP.1 translation stage (TS) in the variable arm controls the pump–probe delay. The beams are focused into the alignment tube using a 50-cm focal-length lens (L1) and subsequently collimated by lens L2. Depending on the measurement, the detection system incorporates an 800-nm bandpass filter (10-nm FWHM), variable neutral-density filter, Ophir PD300-SH power sensor with Vega power meter, Ocean Optics USB4000 spectrometer coupled to a custom-built integrating sphere, or Mightex SME-B050-U monochrome CMOS camera.

The longitudinal component of a radially polarized beam also offers the possibility of aligning molecules along the propagation direction. However, for the 50-cm focal-length lens used in the present experiment, our estimate indicates that the longitudinal intensity is approximately $10^{-4}$ of the transverse intensity. Consequently, longitudinal-field-induced molecular alignment is expected to be negligible under the present focusing conditions.

## 3. Experimental Setup

In this work, a femtosecond laser system based on chirped-pulse amplification (Coherent, Legend L-USP-1K) was used for the impulsive molecular alignment experiments. The laser produced approximately 50 fs pulses at a repetition rate of 1 kHz, with a pulse energy of approximately 1 mJ. The experimental arrangement is shown in Fig. 2. Laser radiation from the chirped-pulse amplifier (CPA) was directed into a four-port Michelson interferometer. The variable arm consisted of a translation stage with a step resolution of 100 nm (Newport, MTM150PP), while the stationary arm contained polarization optics, including a half-wave plate (Thorlabs, AHWP10M-580) or a radial–azimuthal polarization converter (Altech, S-waveplate 800 nm). The alignment tube consisted of an 85 cm long stainless-steel vacuum chamber equipped with high-vacuum CF flange viewports designed to accommodate 1-inch-diameter optical windows (Thorlabs, VC22FL). The windows were 5 mm thick and had an antireflection coating covering the wavelength range from 650 to 1050 nm (Thorlabs,

WG41050-B). A lens with a focal length of 50 cm (Thorlabs, LA1380-B) was positioned immediately before the entrance of the alignment tube to focus the laser radiation into the interaction region. The tube was filled with high-purity nitrogen gas (99.9995%, ALPHAGAZ) at a pressure of approximately 4 bar.

Various diagnostic instruments were positioned after the alignment tube to measure and characterize the molecular alignment signal and evaluate different detection techniques. Optical power measurements were performed using an Ophir photodiode sensor head (PD300-SH, 350–1100 nm) connected to a Vega power meter. Spectral measurements were obtained using an Ocean Optics spectrometer (USB4000, 200–1100 nm) coupled to a homemade integrating sphere. The integrating sphere was employed to minimize measurement artifacts arising from spatial or angular chirp in the output beam introduced by the laser system or experimental optics. Spatially resolved measurements were performed using a CMOS camera (Mightex, SME-B050-U) with a 5 MP sensor and a resolution of 2592 × 1944 pixels. The camera was modified by removing its internal infrared-cut filter to permit detection of the 800 nm laser radiation. Polarization-resolved measurements were performed using a polarizing beam-splitting cube (Thorlabs, CCM1-PBS252, 620–1000 nm). Additional diagnostic optics included a variable neutral-density filter (Thorlabs, NFC-100C-4M) and a hard-coated narrowband interference filter (Thorlabs, FBH05800-10) centered at 800 nm, with a full width at half maximum (FWHM) of 10 nm and a transmission greater than 90%.

Data acquisition was fully automated using LabVIEW and MATLAB. A National Instruments (NI) general-purpose interface bus (GPIB) card installed in the control computer provided communication with the Newport translation-stage controller, enabling automated control of the pump–probe delay. The Ophir Vega power meter communicated with the computer through a USB interface, while the Ocean Optics USB4000 spectrometer was controlled via USB using OmniDriver software. To acquire images from the Mightex CMOS camera, the header file supplied with the camera software was modified to provide compatibility with MATLAB's `loadlibrary` interface and the manufacturer's dynamic-link library (DLL). These interfaces enabled automated acquisition of power, spectral, and spatially resolved measurements as a function of pump–probe delay.

## 4. Results

A central challenge in encoding the angular dependence of molecular alignment into the transverse intensity profile of a probe beam arises from the use of a common-path pump–probe geometry. Although this arrangement ensures spatial overlap between the probe and the molecular ensemble aligned by the pump pulse, it also requires separation of the probe radiation from the pump background following the interaction region. Several detection approaches are available, including polarization-resolved measurements of alignment-induced birefringence [12,18,19], alignment-dependent absorption, frequency conversion through second-harmonic generation followed by dichroic separation, and spectral filtering or analysis of the probe radiation modified by the molecular response [20,24–27].

To identify the most suitable detection method for the alignment microscope, we systematically investigate polarization-resolved detection, integrated power measurements, spectral modulation, and CMOS imaging. These measurements allow us to compare the sensitivity of different detection schemes to the molecular alignment parameter $\Delta A(\tau)$ and its temporal derivative $dA/d\tau$, providing a basis for selecting an effective approach to spatially and angularly resolved imaging of rotational wave-packet dynamics.

### *4.1 Birefringence and Polarization Measurements*

The primary objective of our first experiment was to investigate the spectral composition of the molecular alignment signal, during which we also observed a polarization-dependent response.

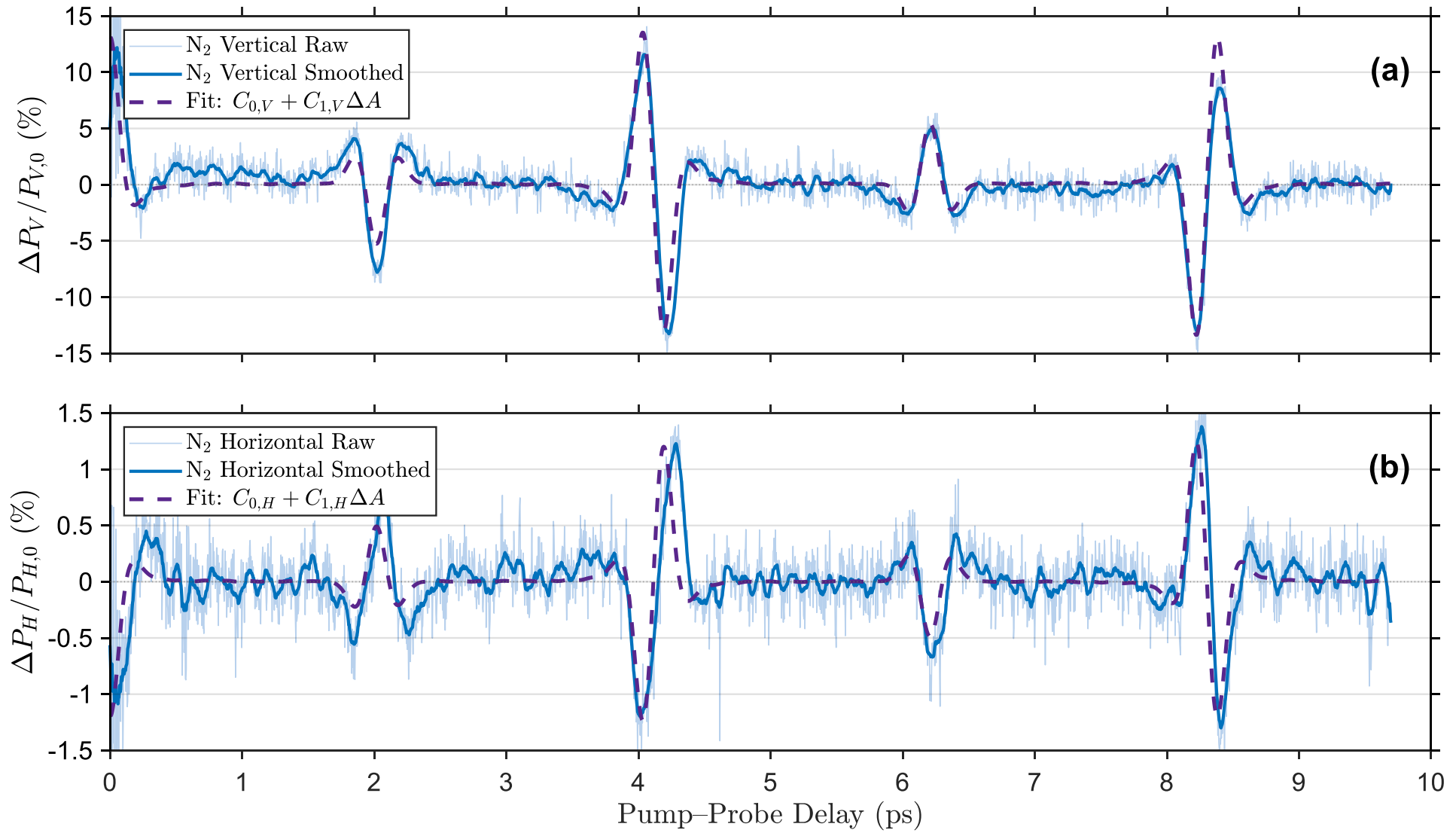

Figure 3: Polarization-resolved measurements of the molecular alignment signal in $N_2$ without spectral filtering. (a) Normalized change in the vertically polarized probe power measured using an Ophir power meter following the polarizing beam splitter (PBS). (b) Corresponding horizontally polarized probe signal obtained by integrating the measured spectrum over 750–850 nm. In both panels, the experimental data are fitted using the molecular alignment parameter, $\Delta A(\tau) = \langle\cos^2\theta\rangle - 1/3$, according to $\Delta P/P_0 = C_0 + C_1\Delta A(\tau)$. Light-blue curves represent the raw experimental data, solid blue curves represent the smoothed data, and dashed purple curves represent the least-squares fits.

The polarization-dependent response of impulsively aligned $N_2$ molecules using a horizontally polarized pump (H-pol) and a probe polarized at 45° (D-pol), with a polarizing beam splitter (PBS) positioned after the alignment tube, as shown in Fig. 2. The vertically polarized output was measured using an Ophir PD300-SH power meter connected to a Vega II readout, while the horizontally polarized output was directed to a homemade integrating sphere coupled to a USB4000 spectrometer. The spectrometer provided additional information on the spectral composition of the alignment signal. For the polarization measurements shown in Fig. 3(b), the horizontal spectrum was integrated over 750–850 nm. Subsequent analysis of the delay-dependent spectra using a software-based 10-nm bandpass filter centered at 800 nm revealed enhanced alignment-dependent intensity modulation, motivating the use of a physical bandpass filter in Sec. 4.2. Both polarization measurements were recorded over approximately 10 ps of pump–probe delay, as shown in Fig. 3.

In addition to our spectral investigations, we observed approximately complementary intensity modulations in the measured horizontal and vertical polarization components. In both panels, the light-blue curves represent the raw data, the solid blue curves represent the smoothed data, and the dashed orange curves represent least-squares fits to the theoretical molecular alignment signal $\Delta A(\tau)$. The complementary behavior suggests that molecular birefringence alters the probe polarization. However, a simple polarization-rotation model predicts complimentary intensities proportional to the square of the molecular alignment signal $\Delta I_H(\tau) \propto -A(\tau)^2$ and $\Delta I_V(\tau) \propto +A(\tau)^2$. Because these quadratic results do not change sign with, they cannot independently reproduce the approximately linear, sign-changing response observed experimentally. We therefore consider the alignment-dependent absorption model introduced in Sec. 2.2, which incorporates the imaginary component of the complex refractive index. The corresponding expressions are presented in Eqs. (12) and (14).

To quantitatively evaluate the relationship between the polarization-resolved signals and the molecular alignment dynamics, the raw, baseline-corrected experimental data were independently fitted using the theoretical alignment response, $\Delta A(\tau) = \langle\cos^2\theta\rangle - 1/3$, according to the linear model $\Delta P / P_0 = C_0 + C_1 \Delta A(\tau)$. The vertical polarization component exhibited good agreement with the theoretical alignment dynamics, yielding $R^2 = 0.720$, whereas the horizontal component showed a weaker correspondence, with $R^2 = 0.428$. The fitted alignment coefficients were $C_{1,V} = 380.27 \pm 8.62$ and $C_{1,H} = -34.91 \pm 1.47$, where the uncertainties represent the half-widths of conventional 95% confidence intervals. The opposite signs of the fitted coefficients indicate complementary alignment-dependent responses in the two polarization components, with the vertical channel exhibiting approximately 10.9 times greater fractional sensitivity to molecular alignment. The RMS contributions of the alignment term were 2.870% and 0.264% for the vertical and horizontal components, respectively. These results indicate that molecular alignment is the dominant contribution to the observed polarization-resolved intensity modulation, particularly in the vertical channel, while the remaining discrepancies suggest the presence of additional optical mechanisms or experimental effects.

The polarization-resolved measurements also reveal a substantial difference in detection sensitivity. The vertical channel exhibited a peak-to-peak modulation of 28.94%, corresponding to an amplitude of 14.47%, with an RMS noise level of 0.961%. This produced a signal-to-noise ratio (SNR) of 15.06 (23.56 dB). In contrast, the horizontal channel exhibited a peak-to-peak modulation of 2.93%, corresponding to an amplitude of 1.47%, with an RMS noise level of 0.230%. Its corresponding SNR was 6.38 (16.09 dB). Thus, the vertical channel exhibited approximately 9.9 times greater fractional modulation and a 2.36-fold higher SNR than the horizontal channel. These results demonstrate the strong dependence of the detected molecular response on the analyzed polarization component and highlight the importance of polarization-resolved detection for optimizing measurement sensitivity.

*4.2 Absorption and molecular alignment cross phase modulation MA-XSPM*

To further investigate the observed alignment-dependent intensity modulation, the PBS, power meter, and spectrometer were removed from the detection arrangement, and collection optics and a power meter were positioned directly after the alignment tube. Inspection of the spectral data in Fig. 3(b) indicated that integration over the wavelength range 795–805 nm produced a pronounced molecular alignment signal. Consequently, a hard-coated bandpass filter centered at 800 nm with a 10 nm FWHM was introduced to investigate the effect of spectral filtering. The power measurements obtained with and without the bandpass filter are shown in Fig. 4. Without spectral filtering, the measured intensity modulation closely follows the theoretical degree of molecular alignment. However, introducing the 800 nm bandpass filter substantially changes the temporal response, which no longer follows the alignment parameter alone. Inspection of the spectra in Fig. 3(b) reveals significant spectral modifications associated with molecular alignment, which are examined further in Sec. 4.3. To explain the filtered response, we consider molecular-alignment-induced cross-phase modulation (MA-XSPM), in which the pump-induced, time-dependent refractive index produces phase modulation of the probe pulse. As derived in Sec. 2.3, this mechanism can produce spectral shifts that are converted into intensity modulation by the bandpass filter.

To quantitatively investigate the relationship between the measured power modulation and molecular alignment dynamics, the raw, baseline-corrected experimental signals were fitted using the theoretical alignment response, $\Delta A(\tau)$, and its temporal derivative, $dA / d\tau$. Without spectral filtering, the experimental response was fitted using the alignment-only model, $\Delta P / P_0 = C_0 + C_1 \Delta A(\tau)$, yielding $R^2 = 0.542$, with fitted coefficients $C_0 = -0.0757 \pm 0.0280$ and $C_1 = 113.26 \pm 3.75$. This indicates that the unfiltered response is predominantly associated with the degree of molecular alignment. With the 800-nm bandpass filter, the measured response was fitted using the combined model, $\Delta P / P_0 = C_0 + C_1 \Delta A(\tau) + C_2 dA / d\tau$, yielding $R^2 = 0.667$, compared with $R^2 = 0.527$ when only the derivative term was included and $R^2 = 0.107$ when only the alignment term was considered. The fitted coefficients for the filtered response were $C_0 = 0.7192 \pm 0.1210$, $C_1 = -290.94 \pm 16.21$, and $C_2 = -35.24 \pm 0.98$, where the uncertainties represent the half-widths of conventional 95% confidence intervals, the power modulation is expressed in percent, and time is measured in picoseconds. The derivative-dependent contribution exhibited approximately twice the RMS amplitude of the alignment-dependent contribution (4.359% versus 2.176%). These findings indicate that the filtered response is more strongly associated with the temporal derivative (66.7%) of molecular alignment than with the alignment parameter itself (33.3%). This behavior is consistent with alignment-induced spectral modulation, including a possible contribution from molecular-alignment-induced cross-self-phase modulation (MA-XSPM), while retaining a significant component directly proportional to the degree of alignment.

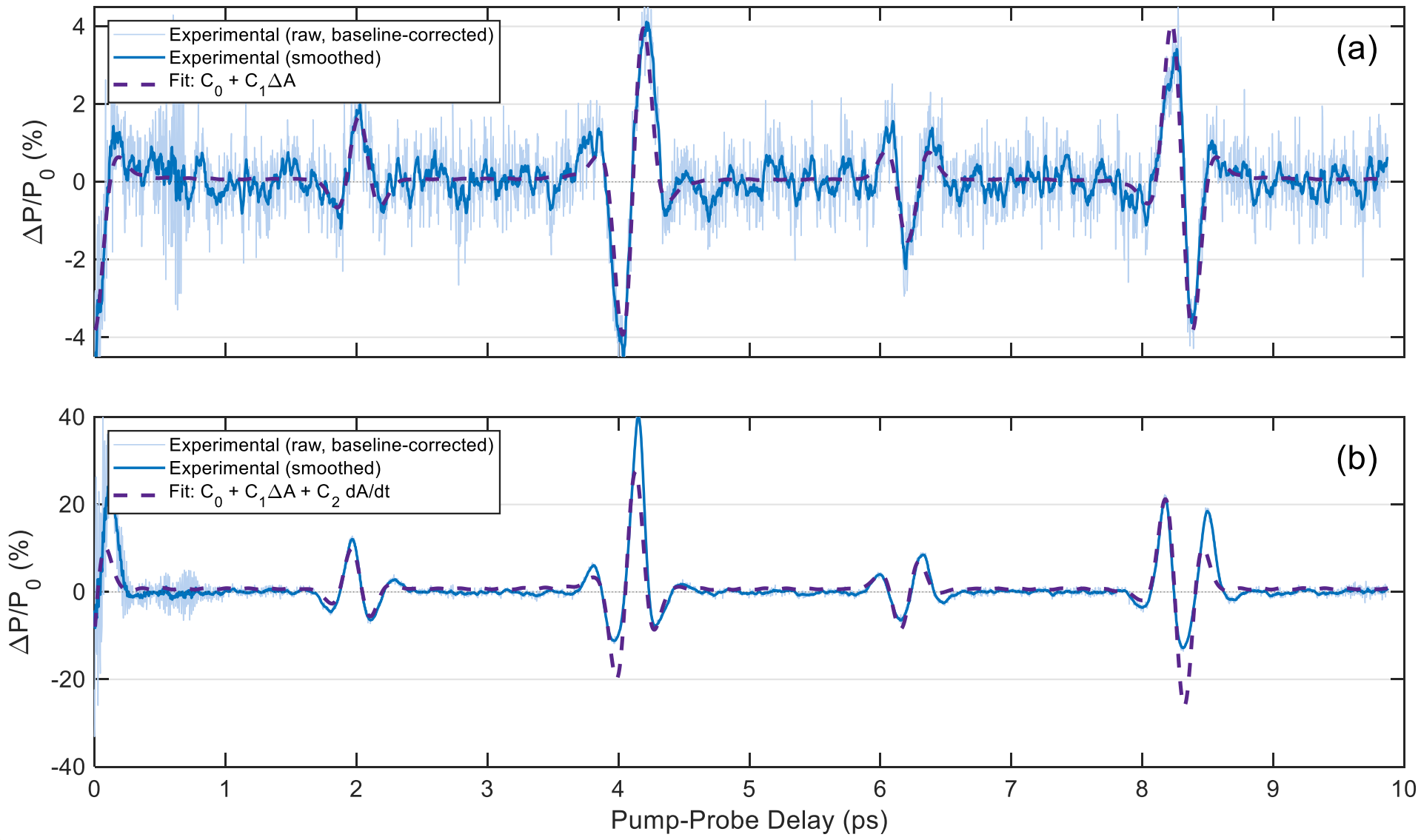

Figure 4: Experimental investigation of absorption-related effects and molecular-alignment-induced cross-phase modulation (MA-XSPM) in impulsively aligned $N_2$. The normalized change in probe power, $\Delta P/P_0$, was measured as a function of pump–probe delay using an Ophir PD300-SH photodetector. (a) Experimental signal recorded without spectral filtering, fitted using only the molecular alignment response, $\Delta A(\tau) = \langle\cos^2\theta\rangle(\tau) - 1/3$. The response is predominantly alignment-dependent, consistent with an absorption-related contribution. The measured signal-to-noise ratio (SNR) is 8.88. (b) Experimental signal recorded with an 800-nm bandpass filter (10-nm FWHM), fitted using a linear combination of $\Delta A(\tau)$ and its temporal derivative, $dA/d\tau$. The filtered response exhibits a substantially stronger derivative-dependent contribution, consistent with spectral redistribution associated with MA-XSPM, while retaining an alignment-dependent component. The measured SNR is 33.52. Light-blue curves represent the raw baseline-corrected experimental data, solid blue curves represent the smoothed data, and dashed curves represent the fitted theoretical responses.

The introduction of the 800-nm spectral filter produced a substantial enhancement in the measured molecular alignment signal. Without spectral filtering, the signal exhibited a peak-to-peak modulation of 11.85%, corresponding to an amplitude of 5.93%, with an RMS noise level of 0.67%. This resulted in a signal-to-noise ratio (SNR) of 8.88 (18.97 dB). With the 800-

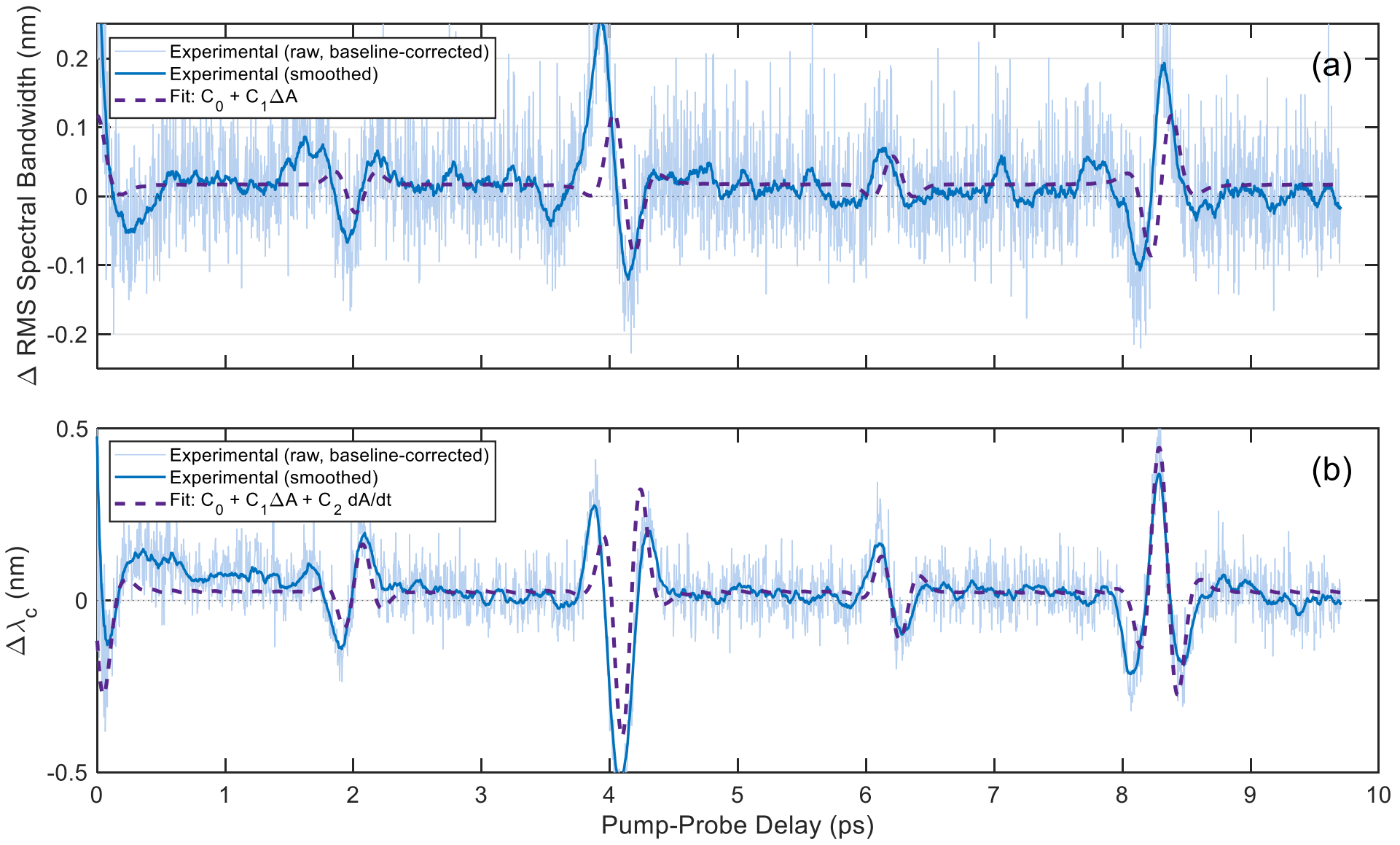

Figure 5: Spectral signatures of impulsive molecular alignment in $N_2$ measured using an Ocean Optics USB4000 spectrometer as a function of pump–probe delay. **(a)** Change in the root-mean-square (RMS) spectral bandwidth, ΔRMS, compared with the calculated degree of molecular alignment, $\Delta\langle\cos^2\theta\rangle = \langle\cos^2\theta\rangle - 1/3$. The experimental bandwidth modulation exhibits temporal features consistent with the predicted alignment dynamics, with a signal-to-noise ratio (SNR) of 5.35. **(b)** Shift in the spectral center wavelength, $\Delta\lambda_c$, compared with the calculated time derivative of molecular alignment, $d\langle\cos^2\theta\rangle/d\tau$. The measured wavelength shift closely follows the temporal evolution of the theoretical derivative, with an SNR of 8.33. In both panels, light-blue curves represent the raw experimental data, solid blue curves represent the smoothed data, and dashed orange curves represent the theoretical predictions. Details of the spectral analysis and signal extraction are provided in the text.

nm filter in place, the peak-to-peak modulation increased to 53.38%, corresponding to an amplitude of 26.69%, while the RMS noise increased to 0.80%. Consequently, the SNR increased to 33.52 (30.51 dB), representing an improvement of approximately 3.77 times, or 11.54 dB. The enhancement arises primarily from the approximately 4.50-fold increase in fractional power modulation, which more than compensates for the modest increase in noise.

To further investigate the mechanisms responsible for the filtered intensity modulation, we examine two spectral observables: the RMS spectral bandwidth and the center-wavelength shift. In particular, spectral changes associated with molecular alignment may provide additional insight into the physical mechanisms responsible for the observed intensity modulation. To examine these effects, we investigate two complementary spectral observables: the root-mean-square (RMS) spectral bandwidth and the shift in the spectral center wavelength. These quantities characterize changes in the spectral width and central frequency of the probe pulse, respectively, and provide a means of evaluating the possible role of MA-XSPM and other alignment-dependent optical processes.

*4.3 Spectral Measurements*

To investigate whether the enhanced signal obtained with spectral filtering arises from molecular-alignment-induced cross-phase modulation (MA-XSPM), the power meter was replaced with an integrating sphere coupled to an optical spectrometer. Figure 5 presents the measured changes in (a) RMS spectral bandwidth and (b) center wavelength. The center-wavelength shift exhibits temporal features associated with the derivative of the theoretical molecular alignment signal. To understand how these spectral changes affect the filtered power, we approximate the 800-nm bandpass filter by a top-hat transmission function of width $\delta$:

$$P(\tau) = \int_{\lambda_f - \delta/2}^{\lambda_f + \delta/2} S(\lambda, \tau) d\lambda \, . \tag{28}$$

Assuming that MA-XSPM produces a small rigid spectral translation $S(\lambda,\tau) \approx S_0[\lambda - \Delta\lambda_c(\tau)]$, a second order Taylor expansion gives

$$\frac{\Delta P(\tau)}{P_0} \approx -\frac{S_0(\lambda_+) - S_0(\lambda_-)}{P_0} \Delta\lambda_c(\tau) + \frac{S_0'(\lambda_+) - S_0'(\lambda_-)}{2P_0} [\Delta\lambda_c(\tau)]^2 \, . \tag{29}$$

Here $\lambda_\pm = \lambda_f \pm \delta / 2$ and $P_0$ is the unperturbed transmitted power. Using the result from Eq. (18) $\Delta\lambda_c(\tau) = (\lambda_0 L I_0 \eta K_0 / 2c)(dA / d\tau) = K(dA / d\tau)$, yields

$$\frac{\Delta P(\tau)}{P_0} \approx C_1 \frac{dA}{d\tau} + C_2 \left( \frac{dA}{d\tau} \right)^2 , \tag{30}$$

where $C_1 = -K[S_0(\lambda_+) - S_0(\lambda_-)] / P_0$ and $C_2 = K^2[S_0'(\lambda_+) - S_0'(\lambda_-)] / 2P_0$ . These results demonstrate that spectral filtering converts the MA-XSPM-induced center-wavelength shift into a measurable intensity modulation. The resulting response contains contributions proportional to the first and second powers of the temporal derivative of molecular alignment, with coefficients determined by the spectral intensity and its slope at the filter boundaries.

To quantify the measured spectral response, the raw, baseline-corrected data were fitted using $Y(\tau) = C_0 + C_1 \Delta A(\tau)$ for the RMS bandwidth change and $Y(\tau) = C_0 + C_1 \Delta A(\tau) + C_2 dA / d\tau$ for the center-wavelength shift. The RMS bandwidth exhibited relatively weak agreement with the theoretical response $R^2 = 0.078$, with $C_1 = 2.911 \pm 0.363$ nm. The center-wavelength shift showed stronger agreement $R^2 = 0.409$, with $C_1 = -3.995 \pm 0.448$ nm and $C_2 = 0.555 \pm 0.027$ nm·ps. The corresponding RMS fitting errors were 0.076 nm and 0.093 nm, respectively. The uncertainties represent half-widths of conventional 95% confidence intervals. The fitted center-wavelength response exhibited relative RMS amplitude contributions of 30.38% from molecular alignment and 69.62% from its temporal derivative.

The RMS bandwidth exhibited a peak-to-peak modulation of 0.655 nm, an amplitude of 0.328 nm, and an RMS noise of 0.0613 nm, yielding an SNR of 5.35 (14.56 dB). The center-wavelength shift exhibited a peak-to-peak modulation of 1.025 nm, an amplitude of 0.513 nm, and an RMS noise of 0.0615 nm, yielding an SNR of 8.33 (18.42 dB). These observations are consistent with alignment-dependent spectral modulation, including a possible MA-XSPM contribution, although other optical processes may also influence the measured spectra.

### *4.4 Comparison of Detection Sensitivity*

To evaluate the relative sensitivity of the different measurement techniques, we compared the signal-to-noise ratios (SNRs) obtained from polarization-resolved detection, integrated power measurements, spectral analysis, and CMOS imaging. The results are summarized in Table 1. For consistency, the SNR was calculated as half the peak-to-peak modulation of the unsmoothed, baseline-corrected experimental signal divided by the standard deviation of the baseline noise. A common signal interval of 3.650–4.550 ps and a baseline interval of 9.000–9.800 ps were used, except for the spectral measurements, for which the available baseline interval was 9.003–9.704 ps.

Table 1. Signal-to-noise ratios for nitrogen molecular-alignment measurements using different detection methods.

| Sec. | Measurement | Unit | P-P | Amp. | Noise | SNR | SNR (dB) |
|---|---|---|---|---|---|---|---|
| 4.1 | Polarization - V | % | 28.936 | 14.468 | 0.961 | 15.06 | 23.56 |
| 4.1 | Polarization - H | % | 2.931 | 1.466 | 0.230 | 6.38 | 16.09 |
| 4.2 | Power - no filter | % | 11.851 | 5.925 | 0.667 | 8.88 | 18.97 |
| 4.2 | Power - 800 nm filter | % | 53.384 | 26.692 | 0.796 | 33.52 | 30.51 |
| 4.3 | RMS bandwidth | nm | 0.655 | 0.328 | 0.061 | 5.35 | 14.56 |
| 4.3 | Center wavelength | nm | 1.025 | 0.513 | 0.062 | 8.33 | 18.41 |
| 4.5 | Camera - no filter | % | 13.424 | 6.712 | 0.638 | 10.52 | 20.44 |
| 4.5 | Camera - 800 nm filter | % | 59.254 | 29.627 | 0.958 | 30.91 | 29.80 |

Note. P-P = peak-to-peak; Amp. = P-P/2; noise = raw baseline standard deviation.

Signal: 3.65-4.55 ps. Noise: 9.00-9.80 ps (spectral: 9.003-9.704 ps).

800 nm filter SNR gain: power 3.77x (11.54 dB); camera 2.94x (9.36 dB).

All SNR values use unsmoothed data; displayed values are rounded.

The introduction of the 800-nm bandpass filter substantially improved the sensitivity of both the integrated power and CMOS measurements. For the integrated power measurements, the SNR increased from 8.88 (18.97 dB) without the filter to 33.52 (30.51 dB) with the filter, corresponding to a 3.77-fold improvement (11.54 dB). Similarly, the CMOS measurements exhibited an increase in SNR from 10.52 (20.44 dB) to 30.91 (29.80 dB), representing a 2.94-fold improvement (9.36 dB).

These improvements are consistent with the spectral sensitivity expected from molecular-alignment-induced cross-phase modulation (MA-XSPM). The time-dependent refractive index associated with the rotational wave packet can modulate the probe spectrum, producing intensity variations when transmitted through a bandpass filter. Consequently, the filtered measurements exhibit a stronger contribution associated with the temporal derivative of molecular alignment, whereas the unfiltered measurements predominantly follow the alignment parameter itself. These results demonstrate that spectral filtering effectively enhances the sensitivity of molecular alignment measurements, particularly for spatially resolved CMOS detection.

### *4.5 Camera Measurements*

Because the alignment microscope employs an azimuthally polarized probe beam in a common-path pump–probe arrangement, polarization-based detection is not well suited to the present configuration. Although polarization-resolved measurements exhibit favorable signal-to-noise ratios (SNRs), the continuously varying polarization direction across the transverse beam profile complicates the separation of the probe signal from the pump background. The SNR comparisons presented in Sec. 4.4 indicate that an 800 nm bandpass filter provides a more suitable approach for detecting alignment-dependent probe intensity modulations. Furthermore, because the objective is to encode the angular dependence of molecular alignment into the transverse intensity distribution, a CMOS camera is required to record the spatially resolved optical response.

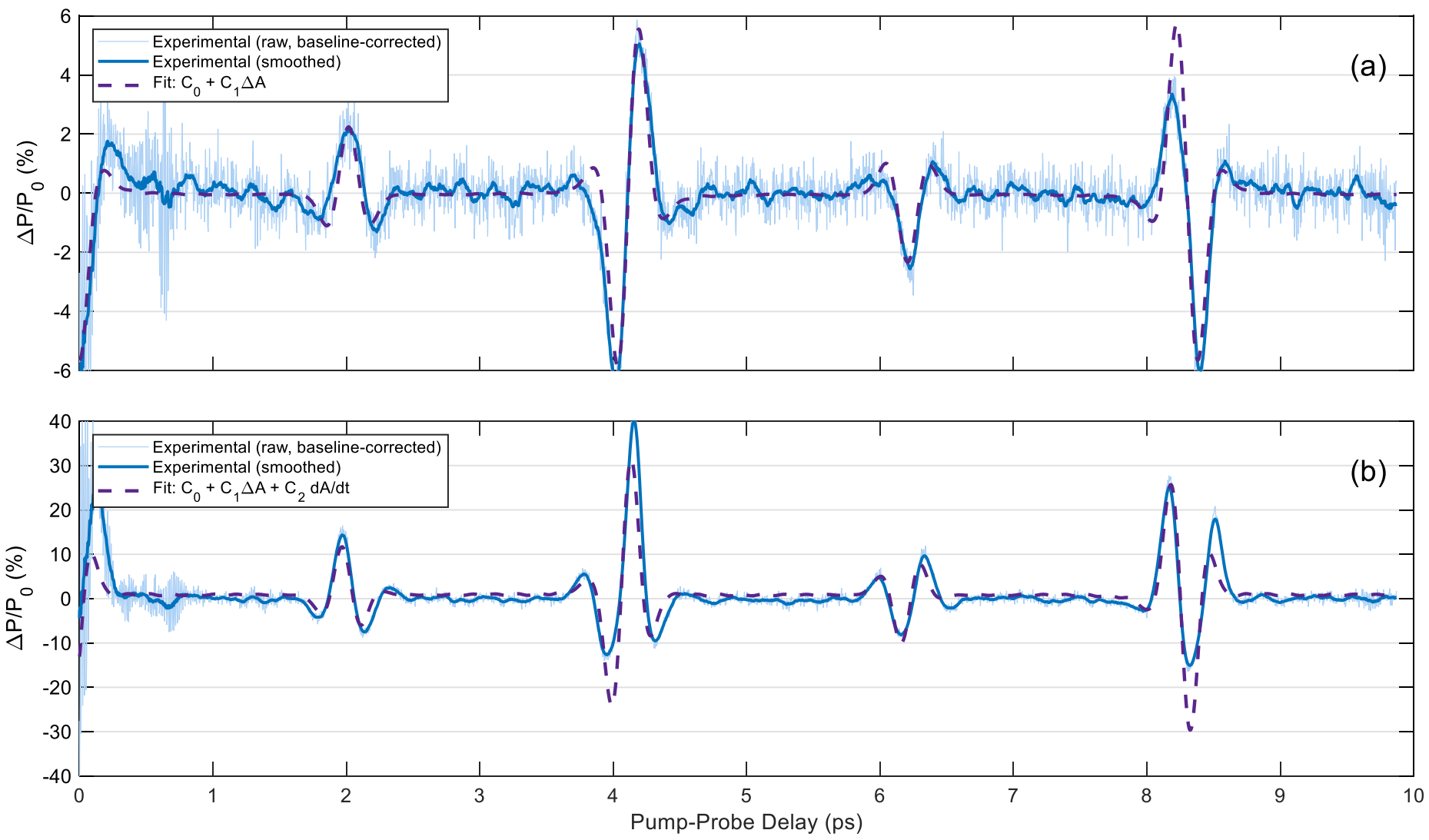

Figure 6: Pump–probe delay dependence of the normalized probe intensity change, $\Delta P/P_0$, obtained by spatially integrating CMOS images of impulsively aligned $N_2$ molecules. (a) Experimental signal recorded without spectral filtering, fitted using only the molecular alignment response, $\Delta A(\tau) = \langle\cos^2\theta\rangle(\tau) - 1/3$. The unfiltered response predominantly follows the degree of molecular alignment, with a signal-to-noise ratio (SNR) of 10.52. (b) Experimental signal recorded with an 800-nm bandpass filter (10-nm FWHM), fitted using a linear combination of $\Delta A(\tau)$ and its temporal derivative, $dA/d\tau$. The filtered response exhibits a stronger derivative-dependent contribution, consistent with spectral redistribution associated with MA-XSPM, while retaining an alignment-dependent component. The measured SNR is 30.91. Light-blue curves represent the raw baseline-corrected experimental data, solid blue curves represent the smoothed data, and dashed purple curves represent the fitted theoretical responses.

To evaluate the suitability of the CMOS camera for quantitative intensity measurements, the camera was initially operated as a power meter by integrating the recorded intensity over the beam profile at each pump–probe delay. Images were acquired using the Mightex camera while scanning the MM3000 translation stage. Figure 6 presents the resulting integrated intensity measurements obtained (a) without spectral filtering and (b) with the 800 nm bandpass filter. The measurements are comparable to those obtained using the conventional power meter in Fig. 4, with the corresponding signal-to-noise characteristics summarized in Table 1. Without spectral filtering, the measured response is predominantly associated with alignment-dependent absorption, whereas the filtered response exhibits a substantial contribution consistent with molecular-alignment-induced cross-phase modulation (MA-XSPM). The bandpass filter significantly enhances the modulation contrast, making spatially resolved imaging of the angular-dependent alignment signal more favorable. Because the CMOS camera captures the entire beam profile, the observed unfiltered intensity modulation is less likely to arise from beam redistribution or deflection than in measurements using the PD300-SH sensor, where spatially varying phase changes could redirect light outside the detector's active area.

To quantitatively evaluate the relationship between the spatially integrated camera signals and molecular alignment dynamics, the raw, baseline-corrected experimental measurements were fitted using the theoretical alignment response, $\Delta A(\tau)$, and its temporal derivative, $dA/d\tau$. Without spectral filtering, the experimental response was fitted using the alignment-only model, $\Delta P / P_0 = C_0 + C_1 \Delta A(\tau)$, yielding $R^2 = 0.656$, with a fitted coefficient $C_1 = -163.09 \pm 4.26$. With the 800-nm bandpass filter, the combined model, $\Delta P / P_0 = C_0 + C_1 \Delta A(\tau) + C_2\, dA / d\tau$, yielded $R^2 = 0.623$, with fitted coefficients $C_1 = -415.17 \pm 20.66$ and $C_2 = -38.42 \pm 1.25$, where the uncertainties represent the half-widths of conventional 95% confidence intervals.

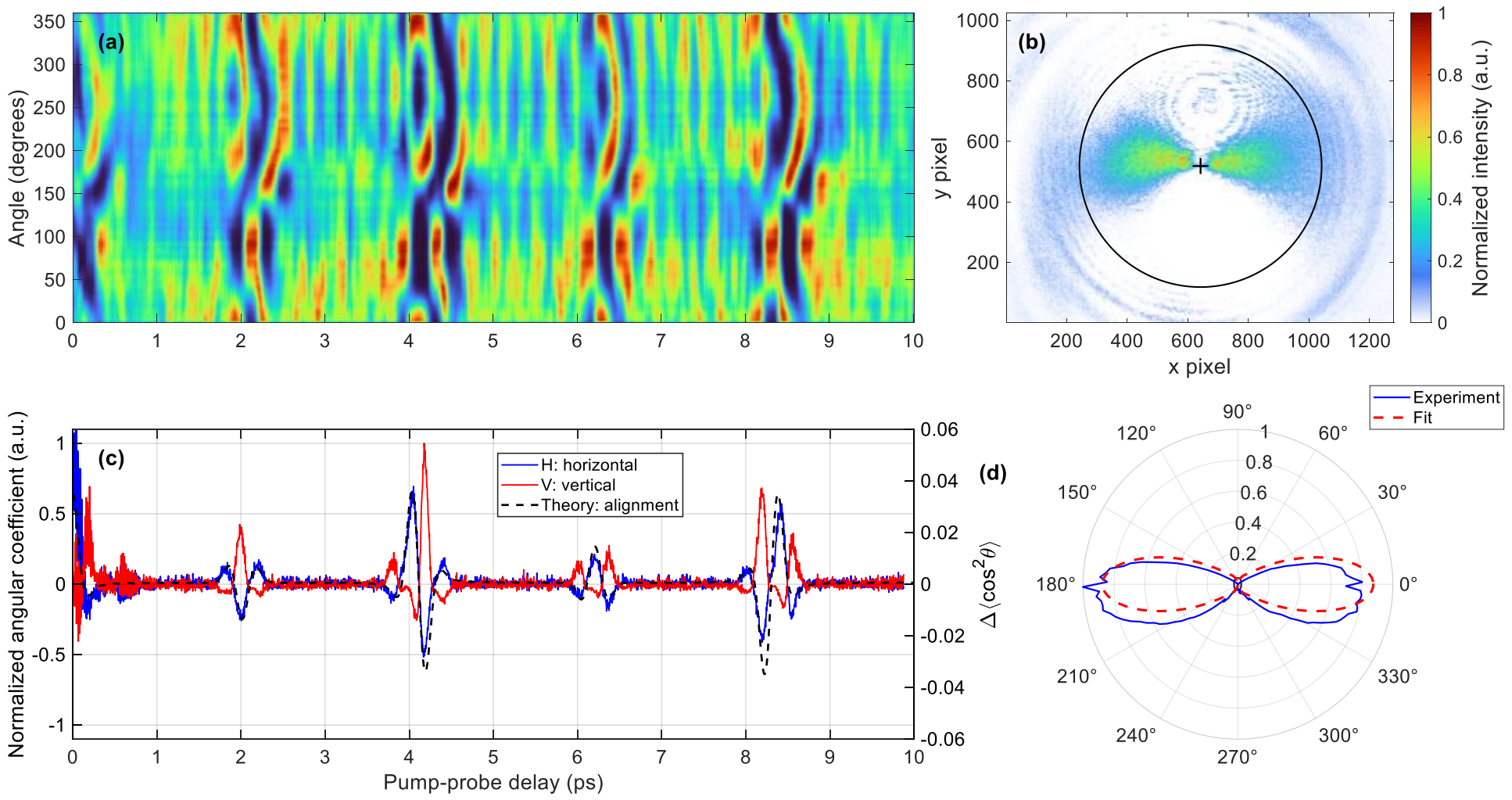

Figure 7: Spatially and angularly resolved optical response of impulsively aligned $N_2$ molecules measured using a CMOS camera. **(a)** Normalized angular intensity distribution as a function of pump–probe delay, revealing pronounced rotational revival structures. **(b)** Representative transverse intensity distribution of the probe beam, with the black circle indicating the region used for angular integration. **(c)** Time-dependent horizontal (H, blue) and vertical (V, red) angular coefficients extracted from the measured intensity distributions, compared with the theoretical degree of molecular alignment, $\Delta\langle\cos^2\theta\rangle$ (black dashed curve, right axis). The approximately complementary H and V responses reveal a pronounced redistribution of the probe intensity between orthogonal angular directions during the rotational revivals. **(d)** Polar representation of the experimentally measured angular distribution (blue solid curve) and its fitted distribution (red dashed curve). Together, these measurements demonstrate that the optical response associated with impulsive molecular alignment exhibits substantial spatial and angular structure, providing a basis for imaging rotational wave-packet dynamics using spatially structured polarization.

The RMS contribution of the alignment-dependent term was 1.224% for the unfiltered measurements. With the 800-nm filter, the RMS contributions of the alignment and derivative terms were 3.115% and 4.752%, corresponding to relative RMS amplitude contributions of 39.60% and 60.40%, respectively. These results indicate that the unfiltered camera response is predominantly associated with the degree of molecular alignment, whereas the filtered response exhibits a substantially greater contribution from its temporal derivative. This behavior is consistent with alignment-induced spectral modulation, including a possible contribution from molecular-alignment-induced cross-self-phase modulation (MA-XSPM), although additional optical mechanisms may also contribute.

The signal-to-noise characteristics of the spatially integrated camera measurements were evaluated using the unsmoothed, baseline-corrected experimental data over a signal window of 3.650–4.550 ps and a noise baseline of 9.000–9.800 ps. Without spectral filtering, the camera signal exhibited a peak-to-peak modulation of 13.42%, corresponding to an amplitude of 6.71%, with an RMS noise level of 0.638%. This yielded a signal-to-noise ratio (SNR) of 10.52 (20.44 dB). With the 800-nm bandpass filter, the peak-to-peak modulation increased to 59.25%, corresponding to an amplitude of 29.63%, while the RMS noise increased to 0.958%. The resulting SNR was 30.92 (29.80 dB), representing a 2.94-fold improvement (9.36 dB) relative to the unfiltered measurements. Although spectral filtering increased the baseline noise, the substantially larger modulation amplitude produced a net improvement in measurement sensitivity. These results demonstrate that spectral filtering enhances the visibility of alignment-dependent optical modulation in the spatially integrated camera measurements.

For the next measurement, we placed the radial polarization convertor in to the path of the probe beam to produce azimuthal polarization. CMOS images were taken for each step of the MM3000 translation stage in increments corresponding to 33.36 fs. Figure 7 shows several analysis of this data. The data in panel (a) were obtained by integrating over radial pie slices having an angular resolution of 2º. The resulting angular distributions were assembled as a function of pump–probe delay to construct the quantum carpet shown in panel (a). Conventionally, such measurements require sequential scanning of the pump–probe delay and polarization angle using a motorized translation stage and a motorized rotation stage equipped with a half-wave plate. Panel (b) is an example CMOS image acquired at $\tau = 4.03\text{ps}$ with an outer integration bound denoted by the circle. Panel (c) was obtained by fitting the vertical slices in panel (a) to $I(\phi) = B + H\left|\cos(\phi)\right|^n + V\left|\sin(\phi)\right|^n$ where $B$ is the isotropic (angle-independent) background, $H$ is the horizontal component, $V$ is the vertical component, $\phi$ is the azimuthal angle, and $n = 5$ was the best fit angular sharpness exponent. As an example, $B$ = 11978.5, $H$ = 118308, $V$ = -12589.4 with a coefficient of determination $R^2 = 0.8759$ . along with the fit coefficient is the plotted the change $\Delta\langle\cos^2\theta\rangle$. The more closely resembles $\langle\cos^2\theta\rangle$ than $d\langle\cos^2\theta\rangle/d\tau$ even though an 800 nm band pass filter was used. This behavior may arise from the combined contributions of alignment-dependent absorption and cross-phase modulation. In particular, the angular dependence of the optical response can produce reduced intensity modulation where the local probe polarization is perpendicular to the molecular alignment axis. Unlike the measurements shown in Figs. 4 and 6, where the pump and probe polarizations are parallel and spatially uniform, the azimuthally polarized probe used here samples different polarization orientations across the beam profile. The blue curve in Fig. 7(d) represents the radially integrated intensity distribution from Fig. 7(b), while the red dashed curve shows the corresponding fit.

The principal experimental demonstration of the alignment microscope is presented in Fig. 8, which shows the spatially resolved optical response of impulsively aligned $N_2$ molecules at selected rotational revival times. The upper two rows display representative CMOS images acquired near the maxima of the theoretical molecular alignment response, $\Delta A(\tau) = \langle\cos^2\theta\rangle - 1/3$, identified by the red markers in the central panel. The corresponding pump–probe delays are indicated above each image. At these delays, the measured intensity distributions exhibit pronounced vertically oriented two-lobed structures, indicating a preferential redistribution of the probe intensity along the vertical direction. This behavior is consistent with the polarization-dependent angular response quantified by the horizontal and vertical angular coefficients in Fig. 7(c).

In contrast, the lower two rows of Fig. 8 show images acquired near the minima of the molecular alignment response, identified by the yellow markers in the central panel. At these delays, the transverse intensity distributions undergo a pronounced reorientation, exhibiting predominantly horizontal two-lobed structures. The transition between vertical and horizontal intensity distributions closely follows the alternating maxima and minima of the calculated molecular alignment response, demonstrating a direct correspondence between the rotational wave-packet dynamics and the spatially resolved optical signal. Importantly, this behavior is observed at multiple rotational revivals, indicating that the angular redistribution is reproducibly associated with the evolving molecular alignment rather than an isolated feature of a particular pump–probe delay.

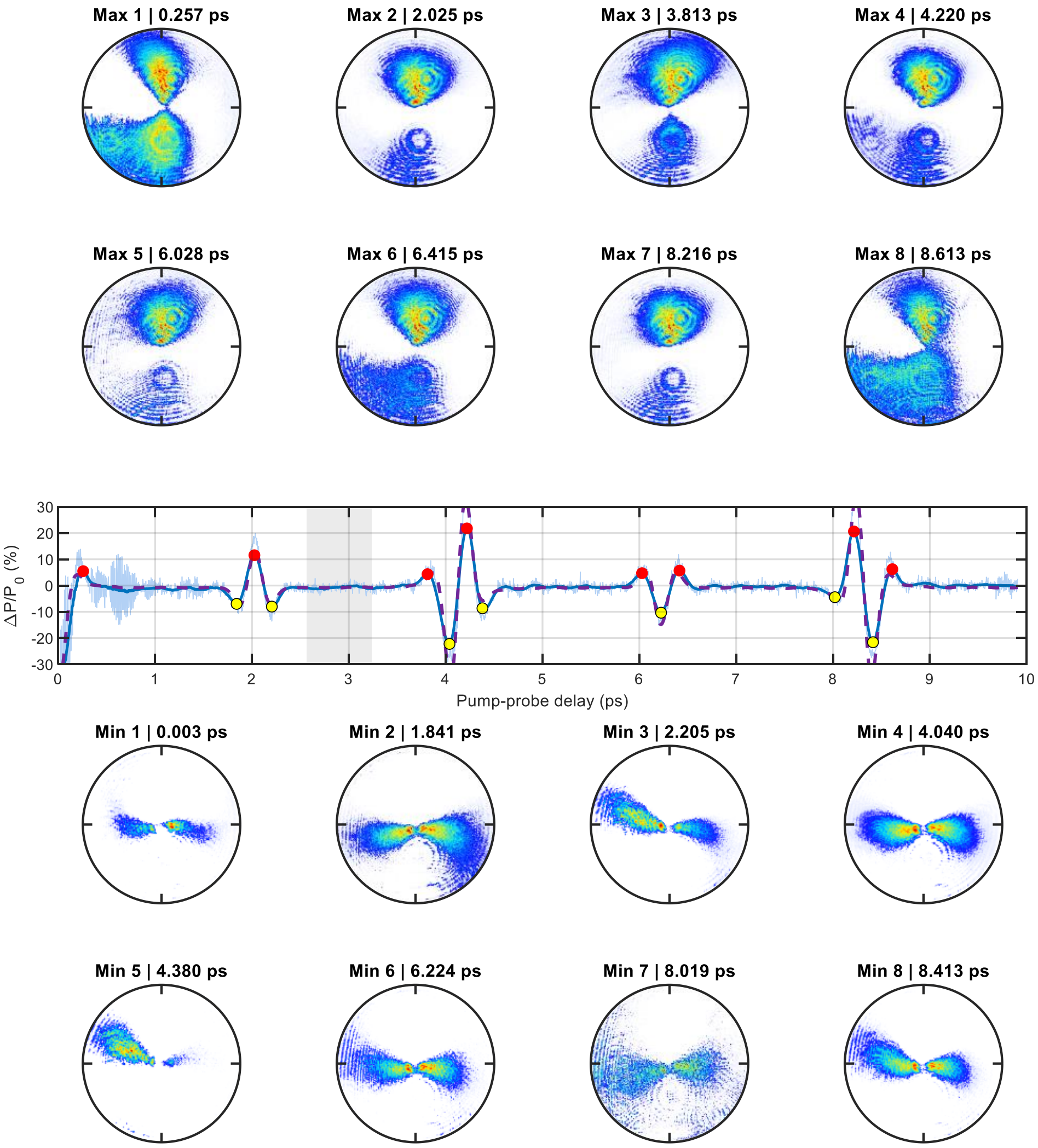

Figure 8: Spatially resolved optical response of impulsively aligned $N_2$ molecules measured using a Mightex SME-B050-U CMOS camera. The upper two rows show transverse probe intensity distributions recorded at selected maxima of the alignment-dependent signal (red markers), while the lower two rows show the corresponding distributions at selected minima (yellow markers). The central panel displays the normalized change in integrated probe intensity, $\Delta P/P_0$ (solid blue curve), as a function of pump–probe delay, together with the theoretical degree of molecular alignment, $\Delta\langle\cos^2\theta\rangle$ (dashed red curve). The pump–probe delay corresponding to each image is indicated above the respective panel. The images reveal pronounced changes in the spatial intensity distribution associated with the rotational wave-packet dynamics of $N_2$. At signal maxima, the intensity is predominantly distributed into two vertically separated lobes, whereas at signal minima, the distribution evolves into two approximately horizontal lobes. These observations demonstrate that the alignment-dependent optical response involves not only changes in the integrated probe intensity but also substantial redistribution of intensity across the transverse beam profile.

The observed reversal of the two-lobed intensity distribution provides a direct spatial signature of molecular alignment dynamics. Unlike conventional polarization-resolved measurements, which require sequential measurements at different analyzer orientations, the alignment microscope simultaneously encodes the angular dependence of the optical response across the transverse probe profile. Consequently, rotational wave-packet dynamics can be visualized through a sequence of CMOS images acquired at successive pump–probe delays without mechanical polarization scanning. These results establish the central capability of the alignment microscope: converting molecular alignment dynamics into a directly observable, spatially structured optical response.

To quantitatively evaluate the relationship between the spatially selected camera signal and molecular alignment dynamics, the raw, baseline-corrected experimental signal was fitted using the combined model $\Delta P / P_0 = C_0 + C_1 \Delta A(\tau) + C_2 dA / d\tau$ , where $\Delta A(\tau) = \langle \cos^2\theta \rangle - 1/3$. The experimental signal in the center panel was obtained by integrating the intensity within a small circular disc directly above the vortex core where the probe polarization is horizontal. The fit exhibited good agreement with the experimental measurements, yielding $R^2 = 0.792$ , with fitted coefficients $C_1 = -956.31 \pm 18.63$ and $C_2 = 16.08 \pm 1.13$ , where the uncertainties represent the half-widths of conventional 95% confidence intervals. The RMS contributions of the alignment and derivative terms were 7.195% and 1.985%, respectively, corresponding to relative RMS amplitude contributions of 78.4% and 21.6%. Thus, the alignment-dependent contribution was approximately 3.62 times greater than the derivative-dependent contribution, indicating that the spatially selected camera signal is predominantly associated with the degree of molecular alignment, with a smaller contribution from its temporal derivative. It should be emphasized that this analysis characterizes the integrated intensity within the selected ROI rather than independently fitting each pixel or the complete two-dimensional intensity distribution. Nevertheless, the agreement between the experimental signal and theoretical alignment dynamics supports the interpretation that the observed angular intensity redistribution during rotational revivals is closely associated with the evolution of the molecular rotational wave-packet.

## 5. Conclusions

We have investigated the optical response of impulsively aligned nitrogen molecules using polarization-resolved detection, integrated power measurements, spectral analysis, and CMOS imaging. The experimental results reveal contributions associated with alignment-dependent absorption and molecular-alignment-induced cross-phase modulation (MA-XSPM). Measurements without spectral filtering predominantly follow the molecular alignment parameter, $\Delta\langle \cos^2\theta \rangle$, whereas measurements using an 800-nm bandpass filter exhibit enhanced intensity modulation and a stronger dependence on its temporal derivative, $d\langle \cos^2\theta \rangle / d\tau$. These observations are consistent with absorption and MA-XSPM, respectively, and demonstrate the importance of both mechanisms in interpreting the measured optical response. Spectral filtering also substantially improved detection sensitivity, increasing the signal-to-noise ratio of spatially integrated CMOS measurements from 10.52 to 30.91.

The polarization-dependent measurements also raise questions regarding the sign of the imaginary molecular polarizability anisotropy, $\Delta\alpha'' = \alpha''_{\parallel} - \alpha''_{\perp}$ . As shown in Eq. (12), a positive value of $\Delta\alpha''$ produces an intensity modulation opposite in sign to the theoretical molecular alignment signal for probe radiation polarized parallel to the pump. This behavior differs from the absorption anisotropy reported in Ref. [22], which implies $\Delta\alpha'' < 0$ . To investigate this discrepancy, quantum-chemical calculations were performed using Gaussian 16 with the CAM-B3LYP functional and aug-cc-pVTZ basis set. Time-dependent density functional theory (TD-DFT) was employed to calculate 30 singlet excited states of $N_2$ under externally applied static electric fields of 0.0169 a.u. (equivalent in amplitude to an optical field with a peak intensity of approximately $\sim 10^{13}$ W/cm$^2$ ) oriented parallel and perpendicular to the molecular axis. The resulting excitation energies and transition dipole moments were used in a sum-over-states calculation of the imaginary polarizability at 800 nm, assuming Lorentzian broadening with a damping energy of 0.1 eV. The calculations yielded positive anisotropies of $\Delta\alpha'' = 4.33\times10^{-3}$ and $3.72\times10^{-3}$ atomic units for the parallel- and perpendicular-field configurations, respectively. The positive sign was maintained for damping energies ranging from 0.01 to 1.0 eV, consistent with the experimentally observed inversion of the horizontal probe signal relative to the theoretical alignment response.

In summary, we have experimentally realized an alignment microscope capable of spatially and angularly resolved imaging of rotational wave-packet dynamics. By employing a spatially varying polarization field, the angular dependence of molecular alignment is encoded directly into the transverse intensity distribution of the probe beam. CMOS imaging revealed complementary horizontal and vertical intensity distributions that evolve with the rotational revivals, enabling simultaneous acquisition of the angular response at each pump–probe delay without mechanical polarization scanning. This technique establishes a new approach to imaging molecular alignment dynamics and provides a foundation for future investigations of nonlinear optical responses and coherent molecular control.

**Disclosures.** The authors declare no conflicts of interest.

**Data availability.** Data underlying the results presented in this paper are not publicly available at this time but may be obtained from the authors upon reasonable request.